\documentclass[twocolumn]{aa}

\usepackage{graphicx}
\usepackage{txfonts}
\begin{document}
   \title{A search for planets  in the metal-enriched binary 
          HD~219542
          \thanks{Based on observations made with the Italian Telescopio 
                  Nazionale Galileo (TNG) operated on the island of La Palma
                  by the Centro Galileo Galilei of the INAF (Istituto 
                  Nazionale di Astrofisica) at the Spanish Observatorio del 
                  Roque de los Muchachos of the Instituto de Astrofisica 
                  de Canarias, and observations collected at the European 
                  Southern Observatory, Chile, using FEROS spectrograph
                  at 1.5 m telescope (proposal ID: 69.D-0338)}}


   \author{S. Desidera
          \inst{1},
           R.G. Gratton
          \inst{1},
           M. Endl
          \inst{2},
           M. Barbieri
          \inst{3},
           R.U. Claudi
          \inst{1},
           R. Cosentino
          \inst{4,5},
          S. Lucatello
          \inst{1,6},
          F. Marzari
          \inst{7},
          \and
           S. Scuderi
          \inst{3}}

   \authorrunning{S. Desidera et al.}

   \offprints{S. Desidera}

   \institute{INAF -- Osservatorio Astronomico di Padova,  
              Vicolo dell' Osservatorio 5, I-35122, Padova, Italy \\
              \email{desidera,gratton,claudi,lucatello@pd.astro.it}            
         \and 
         McDonald Observatory, The University of Texas at Austin, Austin, 
          TX 78712, USA \\
          \email{mike@astro.as.utexas.edu}    
          \and
         CISAS, c/o Dipartimento di Fisica, Universit\`a di Padova, 
         via Marzolo 8, 35131 Padova, Italy \\
          \email{mauro.barbieri@pd.infn.it}
          \and
         INAF -- Oss. Astrofisico di Catania, Via S.Sofia 78, Catania, Italy\\
          \email{scuderi@ct.astro.it}
          \and
         INAF -- Centro Galileo Galiei, Calle Alvarez de Abreu 70, 38700 
            Santa Cruz de La Palma (TF), Spain \\
          \email{cosentino@tng.iac.es}
         \and 
         Dipartimento di Astronomia -- Universit\'a di Padova, Vicolo
         dell'Osservatorio 2, Padova, Italy 
         \and 
         Dipartimento di Fisica -- Universit\'a di Padova, Via Marzolo 8,
         35131 Padova, Italy \\
          \email{francesco.marzari@pd.infn.it}}

   \date{Received 3 Febraury 2003 / Accepted 21 March 2003}

   \abstract{The components of the wide binary HD~219542 were recently
             found to differ in metallicity by about 0.1 dex (Gratton
             et al. \cite{paper1}). In this paper, we present the
             results of  2 years of high precision radial velocity
             monitoring of these stars performed at the Telecopio Nazionale 
             Galileo (TNG) using the high resolution spectrograph SARG. 
             No indication for radial velocity variations above
             the measurement errors ($\sim 5$ m/s) was found for the
             metal richer component A. This allows us to place upper
             mass-limits for planets around this star. 
             HD~219542~B instead shows a low amplitude
             variation with a 112 day period at a  confidence level
             of $\sim 96-97$\%.
             This might suggest the presence of a Saturn-mass planet, although
             it is still possible that these variations are due
             to moderate activity of the star.
             Tests based on variations of bisectors, stellar magnitude
             and line equivalent widths were inconclusive so far.

   }

   \maketitle
%

\section{Introduction}

The increasing number of extrasolar planets discovered by radial
velocity surveys (over 100, see e.g. 
Butler et al.~\cite{butler02}, Butler et al.~\cite{butler03},
Udry et al.~\cite{udry02}) 
reveals a surprising variety of planetary systems:
giant planets very close to their parent stars, planets in very
eccentric orbits, multiple planets systems, in some cases
showing orbital resonances.
Further dedicated efforts are required for
a detailed understanding of frequency and properties of giant planets as a 
function of some fundamental parameters (mass, chemical composition of 
parent star, dynamical environment).

More than half of the stars in the solar neighborhood are in  
binary systems.
Recent discoveries of planets in binary systems
with different mass ratios and separations indicate that planets might 
form in binaries.
The case of 16 Cyg B is remarkable, showing radial velocity variations 
due to a massive planet orbiting the star at 1.7~AU 
(Cochran et al.~\cite{16cyg}). The very high eccentricity (e=0.68) might be 
induced by dynamical perturbations of 16 Cyg A (Holman et al.~\cite{holman97}; 
Mazeh et al.~\cite{mazeh97}). Even more compelling is the case of $\gamma$ Cep,
that harbours a planet in spite of the presence of a low-mass stellar
companion in a 70-year orbit (Hatzes et al.~\cite{gammacep}).
Zucker \& Mazeh (\cite{zucker02}) noted that planets found 
in binaries have a different
mass-period correlation than planets orbiting single stars, with 
an overabundance of short-period high-mass planets.

Besides the study of dynamical effects (minimum binary separation for
planets formation, perturbations of planetary orbits),
binary systems could be used to explore the origin of the 
connection between high metallicity and occurrence of planetary companions
(Gonzalez et al.~\cite{gonzalez01}, Santos et al.~\cite{santos}).
Two possible explanations were suggested: 
either a high metal content makes it easier to form planets
(and in this case no metallicity difference is expected
between members of a binary system  with and without planetary companions)
or the high metallicity is the result of planets or planetesimal ingestion
(Gonzalez \cite{gonzalez97}). 
It is also possible that both explanations are valid.

   Infall of some planetary material on the star surface is expected
   during the formation and subsequent evolution of a planetary system.
   This may happen during the early phases of the planet formation, 
   or in a more mature stage of the planetary system evolution, with
   the infall of remnant of the protoplanetary disk  or with the ingestion 
   of already formed planets, scattered toward the star by dynamical 
   interactions with other planets in the system, distant 
   stellar companions or passing stars.

   If the material engulfed by the star is depleted in hydrogen and helium,
   the chemical composition of the star can be modified, leading to
   metal rich convective envelopes. This alteration is small or negligible
   for stars in the pre main sequence phase, that have thick convective
   zones.
   Stellar pollution is likely efficient mostly for main sequence stars
   with masses 1.0-1.3 $M_{\odot}$, that have a convective envelope thin 
   enough to allow a significant metallicity enhancement for reasonable 
   quantities of accreted material.

   The search for stellar pollution signatures was prompted by the 
   evidence that the host stars of the extrasolar planets found by
   radial velocity surveys are typically metal--rich.
   The stars with planets were searched for trends in metallicity
   as a function of stellar mass or size of convective envelope
   (Santos et al.~\cite{santos} , Pinsonneault et al.~\cite{pinsonneault}), 
   for anomalous abundance ratios, signatures of accretion of chemically 
   fractionated material (Smith et al.~\cite{smith}), and
   for lithium content anomalies (Ryan \cite{ryan}, Israelian et al. 
   \cite{israelian}, Reddy et al.~\cite{reddy}). Lithium, and
   particularly the $^6$Li isotope is in fact a good diagnostic for accretion 
   since it is destroyed during the stellar lifetime.
   Murray et al. (\cite{murray}) studied instead field main sequence stars 
   with spectroscopically determined iron abundance. They found evidences for
   typical accretion of $0.5~M_{\oplus}$ of iron during the main sequence
   lifetime. 

   However, these results are hardly conclusive because of the lack of
   a suitable reference representing the original, unpolluted abundances.

   Wide binaries represent instead easier targets to search for chemical
   anomalies, because of the comparison immediately provided by
   the two companions. Furthermore, their physical association (i.e. same
   distance) allows the use of differential analysis
   between the components. 
   In fact, some of the most compelling results 
   on the chemical anomalies of the stars with planets were achieved 
   by studying the stars with planets member of 
   binary systems  e.g.  16 Cyg (King et al.~\cite{king}, 
   Laws \& Gonzalez \cite{laws}) and 
   HD 178911 (Zucker et al.~\cite{zucker01}).

   Differential techniques for the determination of stellar abundances 
   are very sensitive for pair of  stars with
   similar temperatures and gravities, i.e. main sequence stars with
   similar masses. 
   This argument was at the basis of our project for the study of 
   a sample of wide binaries, aimed to search for planets (using
   radial velocities) and the chemical composition signatures due to
   the engulfment of planetary material.
   This survey is on-going at TNG (the Italian Telescopio Nazionale
   Galileo, INAF, on La Palma, Canary Islands, Spain), using the high 
   resolution spectrograph SARG (Gratton et al.~\cite{papersarg}).

   The first result of our survey was the discovery of a pair (HD 219542)
   whose components differ in iron content by about 0.1 dex 
   (Gratton et al.~\cite{paper1}, hereinafter Paper I), with hints that 
   the overabundance 
   of the primary is confined to rocky elements, while volatiles have
   comparable abundances. The primary is also lithium rich.
   The most likely explanation of this abundance pattern 
   is the infall of planetary material on the primary, either portions
   or remnants of a circumstellar disk or already formed planets.
   Reddy et al.~(\cite{reddy}) searched for $^{6}$Li in HD~219542~A.
   This isotope  represents a very powerful trace
   of recent accretion. They find $^{6}$Li/$^{7}$Li=$0.03 \pm 0.03$, 
   clearly not conclusive. It must be noted that
   such an amount of $^{6}$Li is nearly the one expected on the
   basis of the observed iron difference.

   In this paper, we present the results of the 2 years radial velocity 
   monitoring of the components of HD 219542 aimed at finding planets around 
   them. In this case, we are searching for planets around stars showing
   evidence of accretion of planetary material. This effort is complementary
   to the searches for abundance anomalies of stars around which planets
   have already been detected, and may  represent an important piece in 
   the understanding of the origin of the observed chemical composition 
   differences.
   
   The paper is organized as follows: in Sect. 2 and 3  the available
   information concerning the stellar and orbital characteristics of the
   HD 219542 binary system are described; 
   in Sect. 4. our high precision radial 
   velocity measurements are presented; in Sect. 5 the significance
   of the variability of HD~219542~B is discussed; in Sect. 6 we discuss the
   possibility that the observed variations are due to activity--related 
   phenomena; in Sect. 7  upper limits to the 
   planets still allowed to exist around HD~219542~A are derived; 
   in Sect. 8 the implications of our result are discussed.


\section{Stellar properties}

HD 219542 (HIP 114914, BD -02 5917, ADS 16642) is a wide pair of main
sequence solar type stars.
The physical associations of the components was confirmed by the 
Hipparcos astrometry.
The main properties of the components of HD 219542 are summarized in 
Table~\ref{t:star_param}.

\begin{table}[h]
   \caption[]{Stellar properties of the components of HD 219542.}
     \label{t:star_param}

       \begin{tabular}{lccc}
         \hline
         \noalign{\smallskip}
         Parameter   &  HD 219542 A &  HD~219542~B  & Ref. \\
         \noalign{\smallskip}
         \hline
         \noalign{\smallskip}

$\alpha$ (2000)          & 23~16~35.24 &  23~16~35.42 &  1   \\
$\delta$ (2000)          & -01~35~11.5 &  -01~35~07.0 &  1   \\
$\mu_{\alpha}$ (mas/yr)  & 163.28 $\pm$ 2.12 & 163.28 $\pm$ 2.12  & 1   \\
$\mu_{\delta}$ (mas/yr)  &  20.97 $\pm$ 1.52 & 20.97 $\pm$ 1.52  & 1   \\
RV     (km/s)            &  -11.1 $\pm$ 0.2 & -10.0 $\pm$ 0.2 & 2 \\
$\pi$  (mas)             & \multicolumn{2}{c}{18.30 $\pm$ 1.97}   & 1   \\
$d$    (pc)              & \multicolumn{2}{c}{54.6 $\pm$ 5.9}  & 1 \\
$U$   (km/s)             & \multicolumn{2}{c}{-40.5 $\pm$ 4.8 }  & 2 \\
$V$   (km/s)             & \multicolumn{2}{c}{-16.4 $\pm$ 3.6 }  & 2 \\
$W$   (km/s)             & \multicolumn{2}{c}{ -4.0 $\pm$ 2.5 }  & 2 \\

 & &  &   \\
$V_{T}$                  & 8.172 $\pm$ 0.019 &  8.591 $\pm$ 0.028 & 3  \\
$(B-V)_{T}$              & 0.705 $\pm$ 0.029 &  0.791 $\pm$ 0.044 & 3   \\
V                        & 8.198 $\pm$ 0.021 &  8.589 $\pm$ 0.021 & 4\\
V-I                      & 0.641 $\pm$ 0.035 &  0.705 $\pm$ 0.037 & 4\\
$H_{p}$                  & 8.312 $\pm$ 0.006 &  8.689 $\pm$ 0.008 & 1  \\
$H_{p}$ scatter          & \multicolumn{2}{c}{0.011$^{\mathrm{a,b}}$} & 1 \\
J                        & 6.992 $\pm$ 0.023 & 7.216 $\pm$ 0.033 & 5 \\
H                        & 6.726 $\pm$ 0.039 & 6.896 $\pm$ 0.045 & 5 \\
K                        & 6.627 $\pm$ 0.033 & 6.830 $\pm$ 0.045 & 5 \\
$b-y$                    & \multicolumn{2}{c}{0.405$^{\mathrm{a}}$  $\pm$ 0.006} & 6 \\
$m_{1}$                  & \multicolumn{2}{c}{0.214$^{\mathrm{a}}$  $\pm$ 0.009} & 6 \\
$c_{1}$                  & \multicolumn{2}{c}{0.367$^{\mathrm{a}}$  $\pm$ 0.004}  & 6 \\
 & & &  \\
ST                       & G2V  &  G7V &  7  \\
$M_{V}$                  & 4.48 $\pm$ 0.23 &    4.90 $\pm$ 0.23  & 8 \\
$T_{eff}$ (K)            & 5989 $\pm$ 50 & 5713 $\pm$ 50 &  8 \\
$\Delta T_{eff}(A-B)$ (K)& \multicolumn{2}{c}{276 $\pm$ 14} &  8 \\

$\log g$                 & 4.37 $\pm$ 0.10 &  4.38 $\pm$ 0.10 &  8 \\
 & & &    \\
$\log R^{'}_{HK}$        & -5.36 $\pm$ 0.13 &  -5.11 $\pm$ 0.13 &  2 \\
$ v \sin i $ (km/s)      & 1.9 $\pm$ 0.1 &  2.4 $\pm$ 0.1 &  2 \\
$ v_{mac} $ (km/s)       & 4.4 $\pm$ 0.3 &  4.8 $\pm$ 0.3 &  2 \\
$L_{X}$ (erg/s)          & \multicolumn{2}{c}{$2.1^{+1.8}_{-1.2}~10^{28~\mathrm{a}}$} & 2 \\
 & & &   \\
${\rm [Fe/H]}$           & +0.29 $\pm$ 0.10 &  +0.20 $\pm$ 0.10 &  8 \\
$\Delta {\rm [Fe/H]}(A-B)$& \multicolumn{2}{c}{0.091 $\pm$ 0.009 } &  8 \\
$\log N_{Li}$            & 2.35 &  $<$1.0  & 8 \\
 & & &   \\
${\rm Mass} (M_{\odot})$ & 1.13 $\pm$ 0.05 &  1.06 $\pm$ 0.05 &  2 \\
${\rm Radius} (R_{\odot})$ & 1.05  &  0.96 &  2 \\

Age  (Gyr)               &\multicolumn{2}{c}{$\sim 2$}  &  2  \\

         \noalign{\smallskip}
         \hline
      \end{tabular}

References: 1 Hipparcos (ESA \cite{hipparcos});
            2 This Paper;
            3 Tycho (ESA \cite{hipparcos});
            4 Cuypers \& Seggewiss (\cite{cuypers99});
            5 2MASS;
            6 Olsen (\cite{olsen});
            7 Corbally \& Garrison (\cite{corbally});
            8 Paper I
    
\begin{list}{}{}
\item[$^{\mathrm{a}}$] A+B
\item[$^{\mathrm{b}}$] See Sect.~\ref{s:hipparcos} for details.
\end{list}
\end{table}

\subsection{Stellar masses and ages}
\label{s:age}

Isochrone fitting can only poorly constrain the age of unevolved main
sequence stars. The position in the color-magnitude diagram 
is best fitted by an age in the range 1-2 Gyr.
The other indirect age indicators like chromosperic activity, 
X-ray emission, lithium, and rotational velocity are discussed
in Sect.~\ref{s:rotation}-\ref{s:activity_indic}-\ref{s:galactic}.

The stellar masses derived using 
the Z=0.03 1 and 2 Gyr isochrones of Girardi et al.~(\cite{girardi2000}) are
1.15-1.12 and 1.07-1.05  for A and B component respectively, with an 
uncertainty of 0.05 $M_{\odot}$ due to the errors in the parallax.
In any case, the derivation of ages and
masses using standard stellar models can be misleading for polluted stars
(as likely HD~219542~A), that should have different composition in the 
radiative and convective zone (Sasselov \cite{sasselov}, Sandquist et
al.~\cite{sandquist02}).

\subsection{Rotational profiles}
\label{s:rotation}

Our template spectra have resolution ($R\sim$ 150000) and S/N 
($\sim$ 200) high enough that the shape
of the Fourier Transform of portions of spectra including suitable
lines can be used to derive rotational and macroturbulent velocities.

Stellar rotational velocities have then been derived for both components of the
system (A and B) using the first zeros of the Fourier Transform, as described
by Gray (\cite{gray}). To this purpose, we used
the lines at 6252.565, 6421.360 and 6439.083~\AA~
measured on the radial velocity template spectra.
Assuming a limb-darkening coefficient of 0.6 (appropriate for the stars 
under consideration), we obtain estimates of $1.9\pm 0.1$\ and
$2.4\pm 0.1$~km/s for the values of $v~\sin i$\ for components A and B
respectively. The value for component A agrees quite well with the value of
1.55-2.24~km/s obtained by Reddy et al.~(\cite{reddy}).

The macroturbolence velocities result 
of $4.4\pm0.3$ and $4.8\pm0.3$ km/s 
for HD 219542 A and B respectively, normal for this type of stars.

It should be noted that there is some degree of degeneracy between
the values for rotational and macroturbulent velocities, that is fully
removed only by using very high S/N spectra.
However, the macroturbolence we obtain is  consistent with the
spectral types of the stars, supporting the adopted values for
rotational velocities.

\subsection{Activity indicators}
\label{s:activity_indic}

\begin{figure}
   \includegraphics[width=9cm]{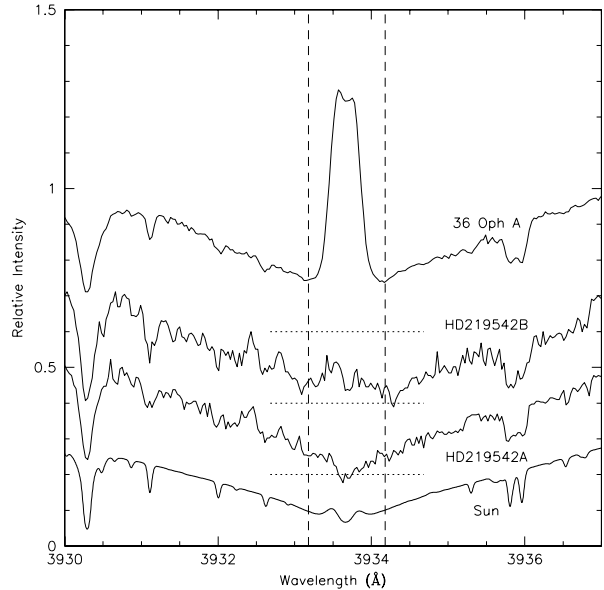}
      \caption{The region near the core of the CaII K line for the Sun
               ($\log R^{'}_{HK}=-4.94$), 
               the components of HD 219542 and the active star 36 Oph A
               ($\log R^{'}_{HK}=-4.58$, Henry et al.~\cite{henry96}).
               The spectra for the other stars were arbitrarily shifted 
               in intensity with respect to the Solar one. The horizontal
               lines represent the zero level for each spectrum. The vertical
               lines show the spectral region used for chromospheric
               activity measurement (the continuum regions are outside
               the plotted range).}
         \label{f:ca2}
   \end{figure}

\begin{figure}
   \includegraphics[width=9cm]{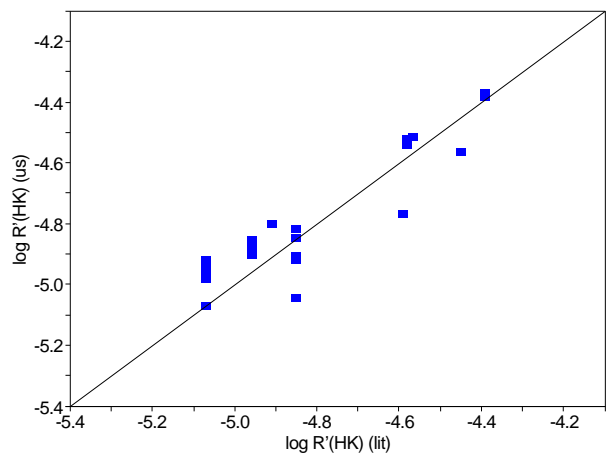}
      \caption{The calibration of $R^{'}_{HK}$ index based on 11 stars
               with literature determinations, observed with the FEROS
               spectrograph during our observing run (August 2002) and 
               instrument commissioning (Kaufer et al.~\cite{kaufer}).}
         \label{f:rhk_cal}
   \end{figure}

The level of chromospheric activity can give some clues about the age of the
system and it is relevant also
for the interpretation of the radial velocity behaviour of the stars
(see Sect.~\ref{s:activity}).
We acquired spectra of both components of HD 219542 (as well as of other 
$\sim$ 40 stars) using the FEROS spectrograph
at the 1.5 m telescope in La Silla in August 2002. Details of observations
will be reported elsewhere. 
The spectra cover the range 3900-9200 \AA, including
the region of chromospheric diagnostics H and K CaII lines.
Fig.~\ref{f:ca2} shows the region of the K line for HD 219542 A and B, compared
to the solar spectrum and to the active star 36 Oph A. 
It appears that both stars have low chromospheric 
activity, with some emission possibly present in  component B
and no evidence for emission in component A.
In order to quantify the activity level, we measured on our spectra the fluxes
within a 1~\AA~band centered on the Ca II K line, and normalized them to 
the values of the fluxes in two narrow {\it pseudocontinuum} bands at 3909 and
3954 \AA. This procedure was repeated for all stars observed with FEROS, as
well as for solar type  stars observed during the FEROS commissioning 
(Kaufer et al.~\cite{kaufer}), whose spectra are
available on the FEROS Consortium
archives\footnote{http://feros.lsw.uni-heidelberg.de/docu/ferosDB/search.html}. 
The resulting flux ratios I(K) were converted
into $R^{'}_{HK}$ values by using 41 observations of 11 stars with 
values of $R^{'}_{HK}$ determined in the literature 
(from Baliunas et al.~\cite{baliunas95}, Henry et al.~\cite{henry96} and
Henry et al.~\cite{henry00}). 
The transformation included a term dependent on (B-V); 
this term accounts for the systematic dependences
of the fluxes in the narrow reference bands on temperature and metal abundance.
The finally-adopted correlation is able to predict $R^{'}_{HK}$ values for the
calibrators with an accuracy of 0.10 dex (see Fig.~\ref{f:rhk_cal}); 
the residual scatter is very
similar to the r.m.s. scatter of $R^{'}_{HK}$ values for the Sun taken at 
different epochs (Baliunas et al.~\cite{baliunas95}), and may be attributed
to the variability of the Ca II emission\footnote{Actually, we expect 
that our values of $R^{'}_{HK}$ contain systematic errors due to stellar 
rotation. Lines of some of our calibrators are indeed significantly 
broadened by rotation; however the values of $R'(HK)$\ we derived for these 
stars do not differ systematically by more than 0.05 dex from the average 
calibration of Figure~\ref{f:rhk_cal}.}.
The values of $R^{'}_{HK}$ for HD~219542 obtained by this procedure are
$-5.36\pm 0.13$\ and $-5.11\pm 0.13$\ for components A and B respectively.
This low level of chromospheric activity is consistent with the results
quoted by Glebocki et al.~(\cite{glebocki}). 
The small rotational velocities we measured are also in agreement with
a low activity level for both the stars. 

The ROSAT satellite detected a source (1RXS J231636.1-013459) that has an
offset of 13 and 18 arcsec with respect HD~219542~A and HD~219542~B 
respectively
(Voges et al.~\cite{rosat}).
Since the position error of the X-ray source is 19 arcsec, 
1RXS J231636.1-013459 is very likely the counterpart of one or both the 
components of HD 219542 (at the high galactic latitude of this star 
the chance of a background object of similar
flux is very small, according to Hasinger et al.~\cite{hasinger}). 
The X-ray luminosity derived using the calibration of
H\"unsch et al.~(\cite{hunsch}) is $L_{X}=2.1^{+1.8}_{-1.2}~10^{28}$~erg/s
(A+B components). Assuming similar contribution by both components,
the X-ray luminosity results about one order of magnitude lower than Hyades 
stars and typical for stars with an age of about 1-2 Gyr 
(Randich \cite{randich}).
Such X-ray luminosity is also in reasonable agreement with the observed
rotational velocities according to the relation by Bouvier (\cite{bouvier}).


\subsection{The galactic orbit and the possible association with
the Hyades}
\label{s:galactic}

\begin{figure}
   \includegraphics[angle=-90,width=9cm]{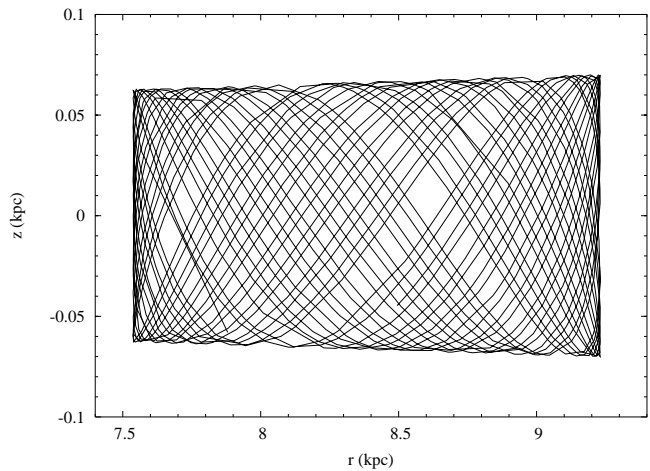}
      \caption{Galactic Orbit of HD 219542 in the meridional plane.}
         \label{f:galactic_orbit}
   \end{figure}

Eggen (\cite{eggen60}) in his search for moving groups 
in the solar neighboorood picked up HD 219542 as a member of the Hyades
moving group. While the existence itself of moving groups is 
debated in the literature (see Taylor \cite{taylor00} and references therein), 
we tried to confirm or
reject this possible  association using updated velocities
and distance for HD 219542 (see Table~\ref{t:star_param}).
The space velocities of the binary are $(U,V,W)=(-40.5,-16.5,-4.0)$.
This is  compatible with the space velocities of the Hyades
given by Perryman et al. (\cite{perryman98}).
The galactic orbit of HD 219542, computed as in Barbieri \& Gratton
(\cite{barbieri}), is represented in Fig~\ref{f:galactic_orbit}.
The orbit ($R_{min}=7.70$~kpc; $R_{max}=9.41$~kpc; $Z_{max}=0.06$~kpc;
$e=0.10$) is normal for a moderately young star.

The distance from the Sun of HD 219542 (54.6 pc) is comparable to that of 
the Hyades. The physical distance of HD 219542 from the center
of the Hyades is about 40 pc.

\begin{figure}
   \includegraphics[width=9cm]{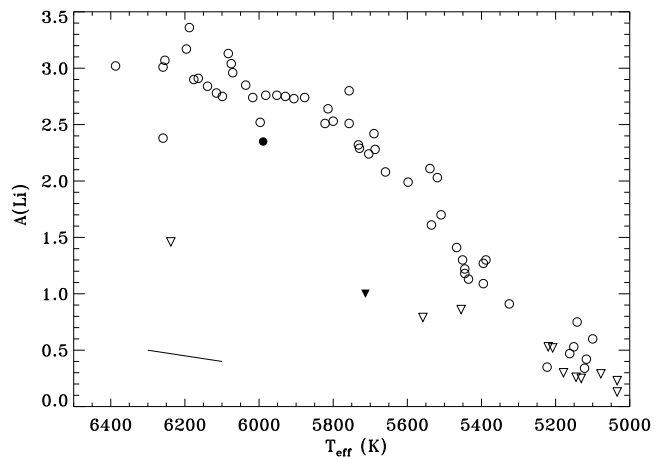}
      \caption{Lithium determinations for Hyades stars from Thorburn
               et al.~(\cite{thorburn}) (empty circles for abundance
               measurements and empty triangles for upper limits) and for the
               components of HD 219542 (HD 219542 A: filled circle;
               HD~219542~B: filled triangle). Only an upper limit
               was derived for HD~219542~B. The line at the lower left 
               corner represents variations of derived lithium content as a 
               function of the adopted effective temperature.
               Even assuming large differences in 
               the temperature scales used by Thorburn et al. (\cite{thorburn})
               and Paper I of $\sim 150-200$ K, it is not possible to
               obtain a lithium content for HD 219542 similar to that of 
               the stars of Hyades of similar temperature.}
         \label{f:lithium}
   \end{figure}

Other physical indicators can be used to shed light on the possible
link between the HD 219542 system and the Hyades:

\begin{itemize}
\item
The metallicity of HD~219542~B (assumed to be unpolluted) is 
[Fe/H]=$+0.20\pm0.10$,
compatible within errors with the cluster metallicity ([Fe/H]=+0.14,
Perryman et al.~\cite{perryman98})\footnote{Note that the [Fe/H] values
given in Paper I may be slightly overestimated, due to the choice
of a very low value for the microturbulent velocity.}
\item
There are no peculiar abundance patterns either in the Hyades stars
or in HD~219542~B
\item
The lithium content of HD~219542~A\footnote{The measurement of lithium
of HD 219542 A of Paper I was confirmed by the independent analysis 
by Reddy et al.~(\cite{reddy}).}
and especially of HD~219542~B
is lower than that of Hyades stars
of similar temperature (Fig.~\ref{f:lithium}) and rather similar to that
of stars in older open clusters like NGC 752 and M67 (Paper I).
This seems not compatible with
the Hyades association, since Hyades show a very low
scatter in lithium abundance at a given temperature.
\item
Isochrone fitting does not have the accuracy for distentangling ages
differences of $\sim$ 0.5 Gyr for unevolved objects like the two 
components of HD 219542
\item
The rotational velocities of the components of HD 219542 are at the
low end of those of Hyades stars of similar spectral type. 
Note that a pole-on orientation for both
components is unlikely, since for binaries with separation
larger than 40 AU the rotation axes do not appear to be aligned
(Hale \cite{hale}).
\item 
The chromospheric activity as measured on our FEROS spectra is much lower than
that of Hyades stars of similar colour 
\item 
The X-ray luminosity derived in Sect.~\ref{s:activity_indic} is about one
order of magnitude lower than that of Hyades stars of similar temperature
\item
None of the 24 Hyades G dwarfs studied by Paulson et al.~(\cite{paulson02})
has a radial velocity jitter as low as HD~219542~A (see Sect.~\ref{s:jitter}).
\end{itemize}

In summary, we conclude that while the kinematic data suggests the 
HD 219542 may be part of the Hyades moving group, the other stellar properties
(lithium, rotation, chromospheric activity, X-ray emission, radial
velocity jitter) seems to indicate a much older age (at least 1-2 Gyr) 
for the system. In particular, isochrone fitting and X-ray emission
point to an age of $\sim 1-2$ Gyr, while the very low 
chromospheric activity suggests an even older age.


\section{Orbital properties of the binary system}
\label{s:bin_orbit}

\subsection{Astrometric Data}

The projected separation of 5.279 arcsec at the distance measured
by Hipparcos (54.6 pc) corresponds to 288 AU. 
Available data can be used to costrain the orbit of the binary.
The first  astrometric measurement  of this pair dates back to 1830
(Aitken \cite{aitken}).
The literature astrometric measurements are listed in Table \ref{t:astrometry}.
Position angles were precessed to 2000.0 using Eq. 5 of Heintz (\cite{heintz}).

\begin{table}
   \caption[]{Relative position of the components of HD 219542.}
     \label{t:astrometry}
      
       \begin{tabular}{ccccccc}
         \hline
         \noalign{\smallskip}
         Epoch      &  $\rho$ & Error & $\theta$ & Error & $\theta_{2000.0}$ & Ref. \\
         \noalign{\smallskip}
         \hline
         \noalign{\smallskip}

1830.51  & 4.56  &       & 26.7  &      & 26.5  & 1 \\
1890.64  & 4.75  &       & 30.6  &      & 30.5  & 1 \\
1909.05  & 4.96  &       & 29.9  &      & 29.8  & 1 \\
1917.32  & 4.96  &       & 30.3  &      & 30.2  & 1 \\
1924.40  & 4.97  &       & 29.2  &      & 29.1  & 1 \\
1953.545 & 5.103 & 0.013 & 30.94 & 0.16 & 30.89 & 2 \\
1960.699 & 5.11  & 0.02  & 30.8  & 0.2  & 30.8  & 3 \\
1976.784 & 5.02  &       & 31.3  &      & 31.3  & 4 \\
1991.25  & 5.279 & 0.004 & 31.53 &      & 31.52 & 5 \\
1991.797 & 5.280 & 0.002 & 31.47 & 0.02 & 31.46 & 6 \\
1994.747 & 5.29  & 0.20  & 32.05 & 0.78 & 32.05 & 7  \\   

         \noalign{\smallskip}
         \hline
      \end{tabular}

References:  
             1 Aitken  (\cite{aitken});
             2 Van Albada - Van Dien (\cite{vanalbada78});
             3 Van Albada - Van Dien (\cite{vanalbada83});
             4 Holden (\cite{holden});
             5 Hipparcos (ESA \cite{hipparcos})
             6 Cuypers \& Seggewiss (\cite{cuypers99});
             7 Popovic \& Pavlovic (\cite{popovic}).

\end{table}

 \begin{figure}
   \includegraphics[width=9cm]{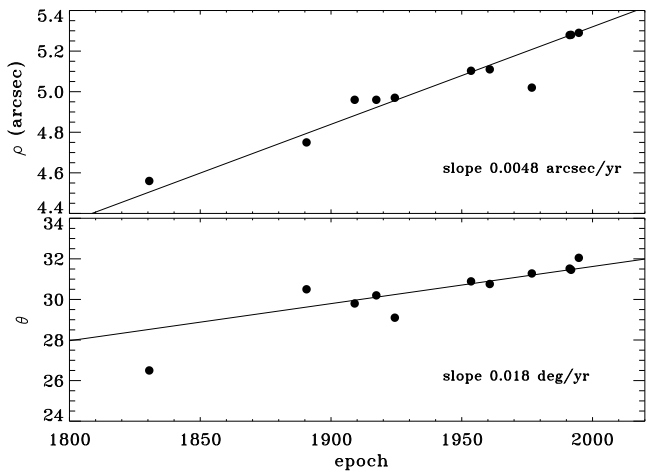}
      \caption{Astrometric measurements of HD 219542.
       Overplotted is the best linear fit to the data.}
         \label{f:rhotheta}
   \end{figure}

Clear trends both in separation and in position angle can be noticed
(Fig.~\ref{f:rhotheta}).
No changes in separation and position angle
were instead detected by Hipparcos because of its limited time baseline.
A linear fitting to the available data\footnote{The measurements were 
weighted according to their errors, when available, and assigning errors
of 0.2 arcsec and 2 deg for $\rho$ and $\theta$ measurements respectively
for those stars without errors bars.}
gives 
$d\rho/dt=0.0048 \pm 0.0003$~arcsec/yr and 
$d\theta/dt=0.018 \pm 0.003$~deg/yr.

\subsection{Radial Velocity Difference}

A further clue to the orbital properties of the system comes from the 
radial velocity differences between
the components, that results $-1.15 \pm 0.02$ km/s (see Sect.~4).
Therefore HD~219542~B is moving toward us with respect to HD~219542~A.

\subsection{Clues on the orbital parameters}

The availability of $\rho$, $\theta$, their time derivatives, the
velocity difference and the parallax means that we know the
relative position of the stars in the plane of the sky ($x$, $y$) and 
all the velocity components $v_{x}$, $v_{y}$, $v_{z}$ (Table \ref{t:posvel}).
The position $z$ along the line of sight is instead unknow, 
so that it is not possible to derive the orbital parameters of the 
visual binary.

\begin{figure}
   \includegraphics[width=9cm]{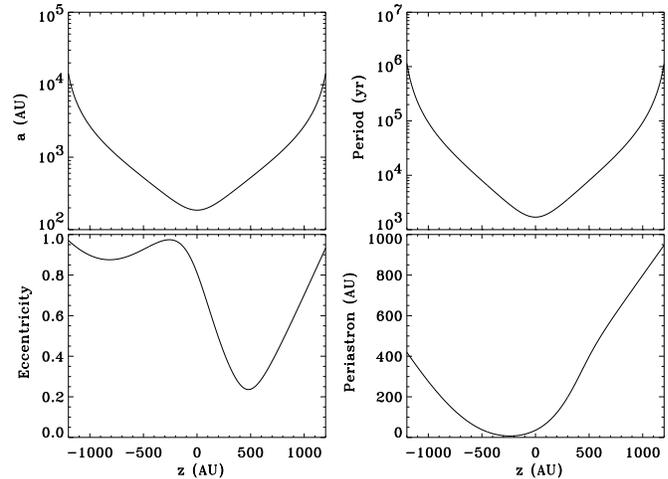}
      \caption{Possible values of semimajor axis, period, eccentricity
                and periastron distance of the orbit of HD 219542
                as a function of the (unknown) separation along
                the line of sight.}
         \label{f:orbit}
   \end{figure}

However, it is possible to constrain the value of $z$ assuming that the
two components are bound and derive the orbital parameters as a function
of $z$. To this aim, we follow  the procedure by 
Hauser \& Marcy (\cite{hauser}). The  allowed separation
along the line of sight results $-1260 < z < 1260$~AU.
The orbital parameters as a function of $z$ are shown in Fig.~\ref{f:orbit}.

\begin{table}
   \caption[]{Relative position and velocities of the components of HD 219542
              in the cartesian plane.}
     \label{t:posvel}
    \centering  
       \begin{tabular}{ccc}
         \hline
         \noalign{\smallskip}
         Parameter      &  Value & Error  \\
         \noalign{\smallskip}
         \hline
         \noalign{\smallskip}

$x$ (AU)       &  246  &  26  \\
$y$ (AU)       &  151  &  16  \\
$z_{min}$ (AU) &-1260  & 280  \\
$z_{max}$ (AU) & 1260  & 280  \\
$v_{x}$ (m/s)  &  840  & 140  \\
$v_{y}$ (m/s)  & 1020  & 110  \\
$v_{z}$ (m/s)  &-1150  &  25  \\

         \noalign{\smallskip}
         \hline
      \end{tabular}

\end{table}

It appears that low eccentricity orbits with periastron of a few hundreds
AU, as well as very eccentric orbits with periastron as small as 10 AU,
are compatible with the available data.
Radial velocity variations due to the binary orbit are below 1 m/s/yr over 
the whole range of $z$, undetectable from our data.

\begin{figure}
   \includegraphics[width=9cm]{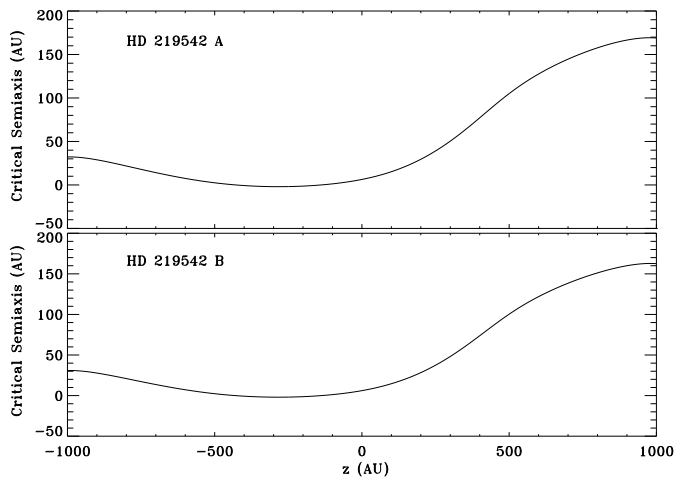}
      \caption{Critical semiaxis for dynamical stability of planets orbiting
               HD~219542~A (upper panel) and HD~219542~B (lower panel) as a 
               function of the current separation along the line of sight.}
         \label{f:acrit}
   \end{figure}

The critical semiaxis for dynamical stability of planets for the family
of orbits described above can be calculated following 
Holman \& Weigert (\cite{holmanweigert}). 
As expected, for the very eccentric orbits occurring for
$-500 < z < 0 $~AU the zone allowed for stable planetary orbits is very 
small,
%
while for the low eccentricy orbits occurring if $z \sim 500$~AU dynamical
stability up to $\sim$ 100 AU is ensured (Fig.~\ref{f:acrit}).

\section{Radial Velocities}

\subsection{Observations and data analysis}

The components of HD 219542 were observed as part of our ongoing planet
search program using SARG at TNG. 
The goals of the project are the study of
the frequency of planets around stars in wide binaries with atypical
separation 200-300 AU, looking for the dynamical effects due to the
stellar companion, and the search for chemical differences between
the components (see Paper I). A recent report is presented by 
Gratton et al.~(\cite{washington}).
The survey as well as  the observational and data analysis procedure will be
described in detail in a forthcoming paper (Desidera et al., in preparation). 
Here we briefly summarize the most relevant points.

The observations used for the radial velocity determinations were
acquired with the SARG spectrograph using the Yellow Grism, that covers
the spectral range 4600-7900~\AA~without gaps, and the 0.25 arcsec slit.
The resulting resolution is  R=150000 (2 pixels sampling).
The iodine cell was inserted into the optical path, superimposing a dense 
forest of absorption lines used as reference spectrum for 
the radial velocity determination.
The spectral ranges 4600-5000~\AA~and 6200-7900~\AA~are free from the
contamination of iodine lines and can be used to monitor
the changes of  the line profile and equivalent width
(see Sect~\ref{s:bisector}-~\ref{s:ew}). Only the blue part of the spectra
is used in this paper.
The insertion of a suitable filter avoids any grism second order
contamination in the red extreme of the spectra.
Exposure times were fixed at 15 minutes, to reduce the  errors in barycentric 
correction caused by the lack of knowledge of the exact flux mid time of the
exposure. 
During the observations, the slit was usually oriented perpendicularly to
the separation of the components
to minimize the contamination of the spectra by the companion.
The rather large separation of the components of HD 219542 (5.3 arcsec)
makes the contamination issue not important for the case presented here.

The data reduction was performed using IRAF\footnote{IRAF is distributed by 
the National Optical Observatory, which is operated by the Association
of Universities for Research in Astronomy, Inc., under contract with the
National Science Fundation}.
 
The radial velocity analysis was performed using the AUSTRAL code developed 
by Endl et al. (\cite{austral}). Each star+iodine spectrum was modeled
using a pure iodine template of the SARG cell obtained with the Kitt Peak
FTS (Desidera \cite{thesis}) and a high quality pure stellar template 
(i.e. without the iodine cell in the path) obtained with SARG, properly
deconvolved for the instrument profile.
In the modeling of the star+iodine spectrum, the best Doppler shift
of the stellar spectrum relative to its template is found and the instrument 
profile was reconstructed.
Since the instrument profile changes along a spectral order, the spectrum
was divided in chunks approximately 100 pixels long ($\sim$2 \AA), each one 
modeled independently. About 400 chunks were used in the final analysis.
Internal errors of velocities are calculated as the errors of the mean
of the chunks. We find that averaging the velocities obtained with several 
different chunk lenghts (in the range 90-120 pixels for solar type stars)
improves the quality of the results. Changing the chunk length means 
to slightly reduce the impact of noise due to the division of the spectrum
in discrete pieces.

The parameters for the instrument profile modeling were instead
kept fixed on values fine-tuned using fast rotating featureless stars,
the constant radial velocity star $\tau$ Ceti and the planet-bearing stars 
51 Peg and $\rho$ CrB.
A long term precision of about 3 m/s was achieved for these stars.
For the much fainter (V=8-9) stars of the binary sample the errors 
are typically about 5 m/s, due to the lower S/N of the spectra.
For the components of HD 219542, the typical S/N we achieved is 80-100.
The stellar templates (used also in Paper I for the abundance analysis)
have instead S/N above 200.

\subsection{Differential Radial Velocities}
\label{s:diff_rv}

More than 20 radial velocities measurements of the components of HD 219542
were obtained, spanning the period October 2000 - November 2002
(Tables~\ref{t:hd219542a}-\ref{t:hd219542b}).
On a few nights, we obtained two consecutive spectra for the targets.
In these cases, we computed a weighted average of the velocities. 

\begin{figure}
   \includegraphics[width=9cm]{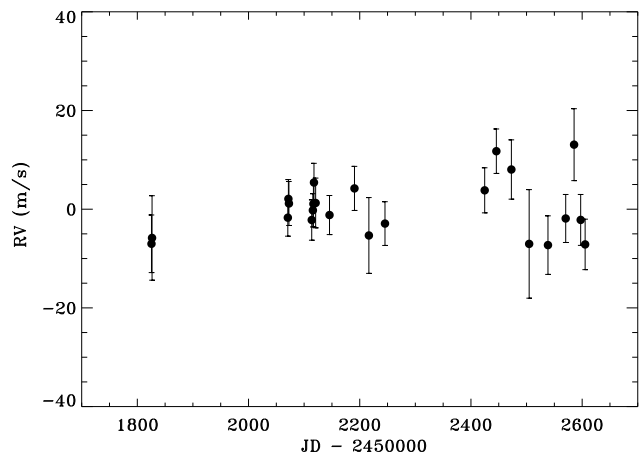}
      \caption{Differential radial velocities of HD~219542~A.
               The dispersion of the velocities is 5.8 m/s, 
               compatible with measurement errors.}
         \label{f:hd219542a}
   \end{figure}

The time series of velocities of HD~219542~A is shown in 
Fig.~\ref{f:hd219542a}.
The dispersion of velocities is 5.8 m/s, only slightly larger than the mean
internal errors (5.6 m/s) determined by the chunk-to-chunk internal
scatter in the velocities\footnote{Comparison between pairs of observations
obtained consecutively in both components suggest that the errors given by 
this procedure may be overestimated by about 25\%, possibly because
individual chunks consistently deviates by similar amounts in spectra
acquired on the same night.}. 
Statistical analysis (see Sect.~\ref{s:statistic}) does not
reveal significant periodicities.
Long term trends are also not apparent from the data.
Therefore we can rule out both short-period Jupiter-mass planets as well
as  massive companions in external orbits.
The upper limits on the giant planets around HD~219542~A still
compatible with  the data are discussed in Sect.~\ref{s:noplanet_a}.

\begin{table}
   \caption[]{Differential radial velocities of HD~219542~A}
     \label{t:hd219542a}
       \centering
       \begin{tabular}{ccc}
         \hline
         \noalign{\smallskip}
         JD -2450000  &  RV & error  \\
         \noalign{\smallskip}
         \hline
         \noalign{\smallskip}

       1825.52  &      -7.2  &     5.8  \\
       1826.59  &      -6.0  &     8.6  \\
       2070.70  &      -1.9  &     3.8  \\
       2071.70  &       1.9  &     3.9  \\
       2072.68  &       1.0  &     4.5  \\
       2113.69  &      -2.4  &     4.1  \\
       2115.67  &      -3.8  &     4.5  \\
       2115.71  &       4.0  &     5.1  \\
       2116.56  &       5.8  &     8.2  \\
       2116.70  &      -1.4  &     5.6  \\ 
       2117.68  &       5.2  &     3.9  \\
       2120.71  &       1.1  &     5.0  \\
       2145.65  &      -1.4  &     4.0  \\
       2190.53  &       4.0  &     4.5  \\
       2216.46  &      -5.5  &     7.7  \\
       2245.42  &      -3.1  &     4.4  \\ 
       2424.70  &       3.7  &     4.6  \\
       2445.70  &      11.6  &     4.5  \\
       2472.70  &       7.9  &     6.0  \\
       2504.67  &      -7.2  &    11.0  \\
       2538.50  &      -7.5  &     5.9  \\
       2570.40  &      -2.1  &     4.9  \\
       2585.46  &      12.9  &     7.3  \\
       2597.36  &      -2.3  &     5.2  \\
       2605.35  &      -7.3  &     5.1  \\

         \noalign{\smallskip}
         \hline
      \end{tabular}

\end{table}

The velocities of HD~219542~B show instead a dispersion (9.2 m/s) larger 
than the measurement errors (5.1 m/s).
Since the two components were usually observed consecutively, we can exclude
the occurrence of systematic errors due to uncorrected instrumental shifts
or barycentric correction as possible causes of the observed variations.

\begin{figure}
   \includegraphics[width=9cm]{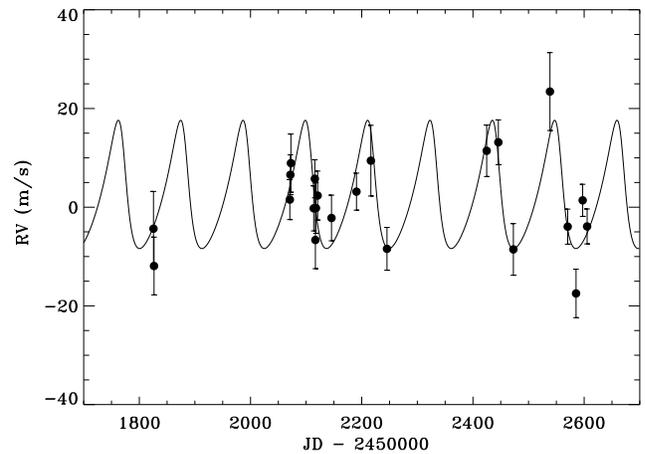}
      \caption{Differential radial velocities of HD~219542~B.
               Overplotted is the best keplerian solution.}
         \label{f:hd219542b}
   \end{figure}

\begin{figure}
   \includegraphics[width=9cm]{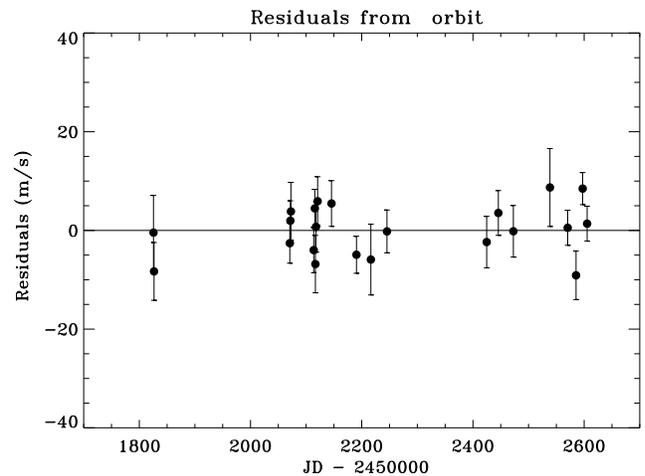}
      \caption{Residuals of the velocities of HD~219542~B from the calculated
               orbit.}
         \label{f:hd219542b_res}
   \end{figure}

\begin{figure}
   \includegraphics[width=9cm]{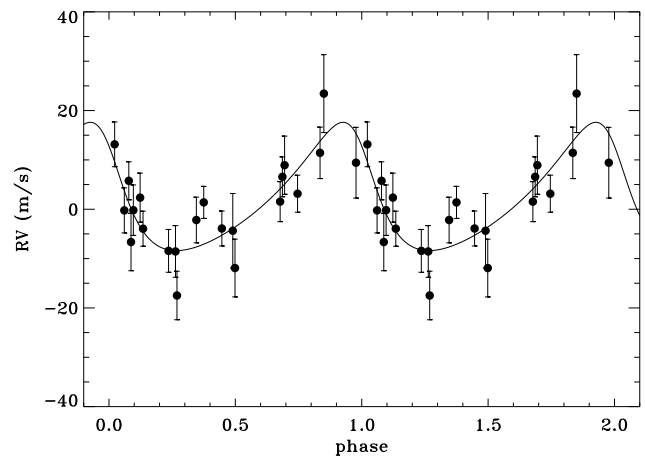}
      \caption{The  velocities of HD~219542~B phased to the 112.1 days period
              of the best orbital solution.}
         \label{f:hd219542b_phase}
   \end{figure}

We performed a period search within the RV-results in the range of $2$
to $1000$ days, using the well-known Lomb-Scargle periodogram 
(Lomb~\cite{lomb}; Scargle~\cite{scargle}).
We found a strong peak at a period of  $112$ days. 
Its significance is discussed in  Sect.~\ref{s:statistic}. 
The best-fit Keplerian orbital solution\footnote{The fitting was
performed using the  program {\it Orbitsolver} (R.E.M. Griffin, priv. comm.)
that adjusts all the elements of the orbit simultaneously in a least-squares
solution (Griffin \& Cornell~\cite{griffin}).}
to the RV-data of HD~219542~B yields a period  P=112.1 days,
semiamplitude K=13.0 m/s and a moderate eccentricity $e \sim 0.32$
(Table \ref{t:planet_orbit}).
The r.m.s. of residuals from the keplerian fit is 5.1 m/s 
(5.9 m/s taking into account the loss of degrees of freedom
introduced by the keplerian fitting).
Fig.~\ref{f:hd219542b} shows the radial velocities of HD~219542~B with
overplotted the orbital solution; Fig.~\ref{f:hd219542b_res}
shows the residuals from the keplerian fit,
and Fig.~\ref{f:hd219542b_phase} the velocities
of HD~219542~B phased to best fitting period.
Fischer et al.~(\cite{fischer01}) noted that for noisy data (and in the
case of undetected further planets in the system) the orbital
fitting produces spuriuos eccentricity enhancements. For this reason,
the derived eccentricity  should be considered with some caution.

\begin{table}
   \caption[]{Differential radial velocities of HD~219542~B}
     \label{t:hd219542b}
       \centering
       \begin{tabular}{ccc}
         \hline
         \noalign{\smallskip}
         JD -2450000  &  RV & error  \\
         \noalign{\smallskip}
         \hline
         \noalign{\smallskip}

       1825.51   &      -4.4       &    7.5       \\
       1826.49   &     -11.9       &    5.9       \\
       2070.71   &       1.5       &    4.1       \\
       2071.71   &       6.6       &    4.0       \\
       2072.69   &       8.9       &    5.9       \\
       2113.71   &      -0.2       &    4.6       \\
       2115.69   &       2.4       &    5.7       \\
       2115.71   &       8.6       &    5.2       \\
       2116.68   &      -6.6       &    5.8       \\  
       2117.70   &      -3.8       &    7.8       \\ 
       2117.71   &       2.5       &    6.8       \\  
       2120.72   &       2.4       &    5.0       \\  
       2145.66   &      -2.2       &    4.6       \\ 
       2190.54   &       3.1       &    3.8       \\ 
       2216.47   &       9.4       &    7.2       \\ 
       2245.43   &      -8.4       &    4.3       \\ 
       2424.71   &      11.4       &    5.2       \\ 
       2445.71   &      13.2       &    4.5       \\ 
       2472.70   &      -8.6       &    5.2       \\
       2538.51   &      23.4       &    7.9       \\
       2570.41   &      -7.8       &    5.2       \\
       2570.42   &      -0.7       &    4.9       \\
       2585.44   &     -16.4       &    7.3       \\
       2585.45   &     -18.4       &    6.7       \\
       2597.33   &      -1.9       &    4.4       \\
       2597.34   &       5.3       &    4.8       \\
       2605.36   &      -4.8       &    4.8       \\
       2605.37   &      -2.8       &    5.3       \\ 

         \noalign{\smallskip}
         \hline
      \end{tabular}
     
\end{table}

\begin{table}[h]
   \caption[]{Possible Orbital Solution for the velocity of HD~219542~B.}
     \label{t:planet_orbit}

       \begin{tabular}{lc}
         \hline
         \noalign{\smallskip}
         Parameter   &  Value \\
         \noalign{\smallskip}
         \hline
         \noalign{\smallskip}

Period (d) &  $112.1 \pm 1.2 $\\
Semi-amplitude (m/s) & $13.0\pm 3.0$\\
Eccentricity & $0.32 \pm 0.15$\\
Longitude of periastron (deg) & $51 \pm $ 33\\
Periastron passage (JD-2450000) & $ 1770.6 \pm 8.0 $ \\
 & \\
$a$ (AU) & 0.46 \\
$m \sin i$ & 0.30 \\
         \noalign{\smallskip}
         \hline
      \end{tabular}
\end{table}

Among the planets found so far, 3 have comparable velocity amplitude
(namely HD 16141b, Marcy et al.~\cite{hd16141}; 47 UMa c, Fischer et 
al.~\cite{47uma}; 55 Cnc c, Marcy et al.~\cite{55cnc}).
Therefore a careful evaluation of the data is required to discuss the
significance of the result (Sect.~\ref{s:statistic}) and the possible
activity origin of the observed variations (Sect.~\ref{s:activity}).
As we will see, the results of these tests are rather inconclusive, so that
the existence of a planet around HD~219542~B still needs 
confirmation\footnote{We cannot then at present confirm at a high
level of confidence the announcement
of a planet that some of us (the italian part of the group) gave as a press
release in November 2002.}.

\subsection{Absolute Radial Velocities}

The iodine cell technique provides velocities relative to an arbitrary
zero point, different for each star. Cross-correlation with a template 
of a star with known absolute radial velocity allows to determine the 
absolute velocity values for both stars.

We used as template 51 Peg, that has an absolute velocity
of $-33.23$~km/s (Nidever et al.~\cite{nidever}).
The velocities result of $V_{A}= -11.1$ km/s
and $V_{B}= -10.0 $ km/s. Corrections for gravitational
redshift and convective blueshift were not included, because 51 Peg has
a colour intermediate between those of HD 219542 A and B. 
It was also not necessary to include a correction for the orbital 
motion due the planet around 51 Peg, because the template of 51 Peg
was acquired by chance with the star close to its center of mass velocity.
The errors of our velocities are likely within 0.2 km/s.
The only literature determination of radial velocity 
is -16.5 km/s for HD~219542~A (Wilson \cite{gcrv}).
The discrepancy is likely explained by the low accuracy of the old data.

For the analysis of the possible binary orbit (Sect.~\ref{s:bin_orbit})
it is important to have a precise measurement of the difference of 
radial velocities between the components. 
Such a difference was measured using the same stellar template
for both stars. Considering the templates of HD 219542 A and HD 219542 B,
the radial velocity difference results $V_{A} - V_{B}= -1200 \pm 10$ m/s.
Such quantity must be corrected for the difference of the gravitational
redshift and convective blueshift between the stars to give the real
radial velocity difference.
We estimate this effect to be about 50 m/s, using the 
relation by Nidever et al.~(\cite{nidever}). Therefore our best
estimate for the radial velocity difference between the components
is  $V_{A} - V_{B}= -1150 \pm 25$ m/s (including further
$\sim 20$ m/s of uncertainty in the gravitational and convective
corrections).


\section{Test of significance of the proposed planet orbit}
\label{s:statistic}

We considered two different tests to evaluate the significance 
of the solution. The first one answers to the question: ``What
is the probability that we would extract a solution corresponding to a 
Keplerian motion induced by a planet better than the solution found
from the real data from a similar analysis of a random sequence
of radial velocities sampled at the same epochs and with the same noise as 
the data ?''

The following procedure was used to answer  this question.
We employed an orbit finder program that scans through the parameter
range for period, amplitude, phase angle, eccentricity and
periastron angles
and determines the quality of the fit for each test-orbit.
We considered two different best fit estimators, minimizing either the r.m.s.
scatter of data or the $\chi^2$ of the distribution, using the
internal errors as estimate of the real errors.
While $\chi^2$ might be expected to be a better fit estimator,
we note that on one side the distribution of individual internal errors is
quite uniform, and on the other internal errors may neglect some source of
scatter present in the data (e.g. activity-induced jitter, 
barycentric correction, etc.). Furthermore, we have some indications
that internal errors are overestimated by $\sim$ 25\% (Sect.~\ref{s:diff_rv}). 
We define as acceptable those solutions whose 
eccentricity is not larger than 0.8,
and there is no phase interval longer than 0.25 without any data.

We verified that our orbit finder program  delivers similar solutions 
to the {\it OrbitSolver} program  using both estimators.
The solutions found by our program have in both cases  period of 111.3 d, 
amplitude of 12.7 m/s, and eccentricity 0.4. 
The r.m.s. of residuals is of 5.8 m/s and the  $\chi^2$ is  104.7 m/s.
In both cases they are very close to the optimized value.

We employ a bootstrap randomization scheme in order to estimate the
significance of the 112 days period.
In this bootstrap method the radial velocities and  errors of individual
nights are randomly
re-distributed along the times of observations.

We ran the program on the generated data sets and we found 
that only in 2 cases out of 1000, random extractions yield
solutions with an r.m.s. smaller than or equal to that given by the
real observations. When using $\chi^2$, the solutions better
than the observed case are 1.2\%.

\begin{figure}
   \includegraphics[angle=-90,width=9cm]{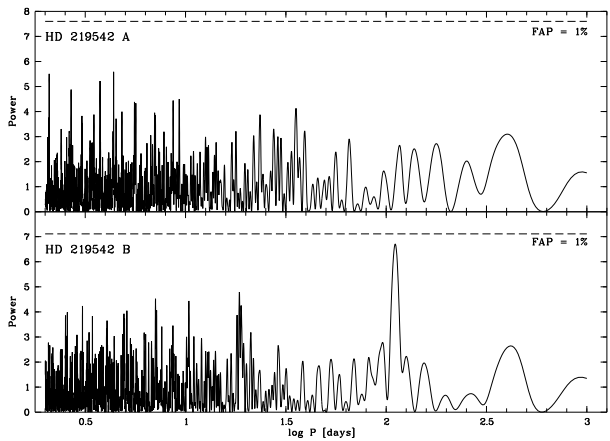}
      \caption{Lomb-Scargle periodogram results for 
               the RV data of component A (upper panel) and
               component B (lower panel). The false-alarm-probability (FAP) 
               levels of $1\%$ are indicated as horizontal dashed lines 
               (see text for details). While there is no signal present for
               HD~219542~A, we find a strong peak at P$=111$ days for 
               HD~219542~B. However, even this signal has a FAP of $3.7\%$ 
               and is thus not highly significant.}
         \label{f:scarglelomb}
   \end{figure}

Second, we tested the probability to have by chance
a power in the periodogram similar to that of the observed data, following  
the bootstrap approach decribed above 
and computing the power spectrum for each new dataset is
generated (K\"urster et al.~\cite{mak97}; Murdoch et al.~\cite{mur}).
For each of the two components
we performed 10000 bootstrap runs. The results are displayed in
Fig.~\ref{f:scarglelomb}. We found
a false-alarm-probability (FAP) of $3.7\%$.
We note that this second approach is more general, 
since it assumes that the signal is given by sine-functions,
independently of its physical origin.
From this more prudent estimate,
we conclude that there is a 3\% probability that our
solution may be due to chance. 

The same tests performed for component A give a rather high
probability (more than 50\%) that a better result can be obtained from a 
random set of radial velocities.

Finally, it should also be remembered that
a basic assumption throughout this procedure is that there is no
correlation between noise on consecutive (or close) nights. However, some 
correlation is expected if noise is dominated by activity. Hence 
this test cannot exclude that the observed velocity variations are due 
to activity rather than to keplerian motion.


\section{Activity or Planet?}
\label{s:activity}

The low amplitude of the velocity variations of HD~219542~B may be
in principle due to orbital motion (i.e. a planet) or to activity related
phenomena. We note that a period of 112 days is too long for a rotational 
modulation (the upper limit to the rotational period from the
observed $v \sin i$ value is $\sim$ 21 days) and
too short for the long term activity cycle typical for solar type stars.
However, it is still possible that the observed variations are due
to some transient short period modulations, not apparent as significant
in the statistical analysis because of the lack of coherence over our
$\sim$ 800 days of observations\footnote{The second peak in 
Fig.~\ref{f:scarglelomb} is at 18.5 days (a reasonable
rotational period for the star). 
While the season 2002 data can be fitted fairly well by a
sinusoidal variation with this
period, the 2001 data cannot.}.
In this case, however, we would have picked up a phase of low activity
during our observations in July 2001, when we obtained 5 observations
over 8 nights, showing a dispersion of 4.5 m/s (nightly averages).

We performed a few tests to check this possibility:
we studied the characteristics of the components of HD 219542 
to infer the expected activity-induced radial velocity jitter, 
we checked the Hipparcos photometry for variability, and we measured
line bisector and equivalent width variability on SARG spectra.

\subsection{Bisector analysis}
\label{s:bisector}

Given the small amplitude of the observed variations in the radial velocity of
HD~219542~B, it is well possible that such variations are not
real, but rather are due to the effects of line asymmetries. 
Such variations could
be in principle related to activity: e.g. in the case of HD~166435, Queloz et
al.~(\cite{queloz01}) found that radial velocity variations with 
semiamplitude as
large as $K=83$~m/s can be attributed to correlated variations of the line
bisectors of $\sim 200$~m/s (peak-to-peak) with the rotational period and
attributable to the effects of star spots. Unfortunately, 
such an analysis is much
more difficult in the present case, because the variations in the bisectors
required to explain the radial velocity variations are very small ($\leq
10$~m/s). While we did not expect that such an accuracy can be achieved using
our data for both components of HD~219542, we however performed an analysis of
the variations of line bisectors, using those orders shortward of 5000~\AA,
where there are not lines due to iodine.

\begin{figure}
   \includegraphics[width=9cm]{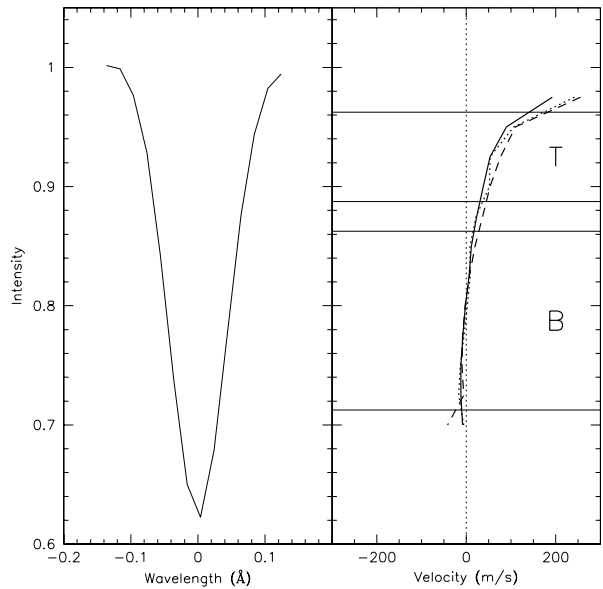}
      \caption{Left panel: Cross Correlation Function for a spectrum of 
               HD~219542~B; right panel: superposition of the the mean 
               line bisectors from the spectra of 51 Peg (dotted line), 
               HD~219542~A (dashed line), and HD~219542~B (solid line). 
               T and B regions are those used in the computation of the 
               bisector span.}
         \label{f:bisector}
   \end{figure}

The technique we used is similar to that devised by 
Queloz et al.~(\cite{queloz01}),
that is an analysis of asymmetries in the Cross Correlation Function (CCF).
The CCF is obtained by cross correlating the observed spectra with a suitable
mask, that is a function that is 1 within very narrow regions (ideally
$\delta$ of Dirac) centered at the
wavelengths of selected spectral lines, and 0 elsewhere. 
This CCF represents an average profile for the
lines in the lists. 
To this purpose we selected about 80 clean spectral lines in
this spectral region, with equivalent widths in the range 10-70~m\AA. 
The left panel of
Fig.~\ref{f:bisector} displays a typical CCF obtained by this technique. 
Typical Full Width Half Maximum (FWHM) of our CCF's are 0.10~\AA: 
this broadening is essentially due to
macroturbulence, since the instrumental profile has a typical FWHM of
0.032~\AA. In analogy with Queloz et al.~(\cite{queloz01}), we quantified line
asymmetry span as the difference $b$\ between the average bisector values in
two regions of the CCF's (from 0.96 to 0.89, and from 0.86 to 0.72 of the line
profile, see Fig.~\ref{f:bisector}).

Our analysis is certainly not optimal: in fact, asymmetries in our CCFs may
arise for reasons different from real asymmetries in the spectral lines (and 
other than photon noise). 
Possible causes include blends with nearby lines, errors
in the wavelength calibration and inhomogeneities in the slit illumination. The
first effect has not a large impact here, 
because it would show up as a constant
pattern, and we are rather interested in bisector variations. 
Also, we selected lines very carefully, and the average line profiles
is very close to the continuum at the base of the line.
We also
note that at least in the case of 51 Peg (for which we performed a similar
analysis using our own spectra, see below), 
the run of the bisector with depth is similar to those
published in the literature (Gray \cite{gray97}) 
in the regions of overlap. 
The runs of the bisectors with depth are very similar for 51 Peg and
the two components of HD~219542, with a trend for the span to slightly
larger in the warmer star (HD~219524~A) and slightly smaller in the cooler
one (HD~219542~B), with 51 Peg (which has intermediate temperature) showing
an intermediate behaviour. This agrees with expectations (Gray~\cite{gray}).
The other
effects are of major concern, and likely are more important in our analysis
than in that by Queloz et al.~(\cite{queloz01}) because slit illumination
is not scrambled by fibers, and because 
our wavelength calibration is not as accurate. On the
other side, it should be also noted that our spectra have a resolution that is 
about three times larger. and this would result in a gain af about one order
of magnitude of the sensitivity of the CCF to real changes in the bisectors.
We also note that all these effects do not affect 
our radial velocities, because they are virtually eliminated in our 
calibration procedure based on the iodine cell lines; however they may
dominate asymmetries in the cross correlation peaks, hiding the effects of
activity-related variations of the bisectors.

In order to quantify the internal errors, we analyzed with the same procedure
a series of 23 spectra of 51 Peg (these spectra have S/N ratios about 2.5 times
larger than those of HD~219542) taken over two years with SARG at TNG.
Spectrum-to-spectrum variations of the bisector span $b$\ are $\sim 10.7$~m/s
(r.m.s.). The scatter is smaller (about 5.6~m/s r.m.s.) for spectra
taken during the same night (generally over a few minutes).
This last value seems then a lower limit for the internal errors in our 
bisector values for this star. In fact, one one hand we expect that 
real variations of the bisectors should show up over longer timescales, and
on the other hand we can not exclude that other errors arise when 
comparing data from different nights, when e.g. a different 
wavelength calibration is used.
Hence, in the case of 51 Peg we may 
set an upper limit to real bisector span variations of about 9~m/s from 
our data.
More stringent upper limit on the bisector span variations for 51 Peg 
was obtained by Hatzes et al.~\cite{hatzes98} ($\sim 7$ m/s).
However, the definition of the bisector span is different in that case, so 
that it is not easy to compare the two results.
Transformation of our  upper limit on bisector to a limit on the jitter 
in radial velocity is not obvious (it depends on the definition of
the bisector span): however, in the case of HD~166435 (for which Queloz
et al.~\cite{queloz01} used a definition similar to the present one)
radial velocity variations are roughly 1.1 times the variations of the 
bisector spans.  If the same rule then applies to our case, 
the upper limit to radial velocity jitter for 51 Peg is
$\sim 10$~m/s: this is about 2.5 times larger than the limit obtained from
photometric (Henry et al.~\cite{henry00}) and equivalent width variations
(see below). However, we consider it as still an interesting upper limit.

\begin{figure}
   \includegraphics[width=9cm]{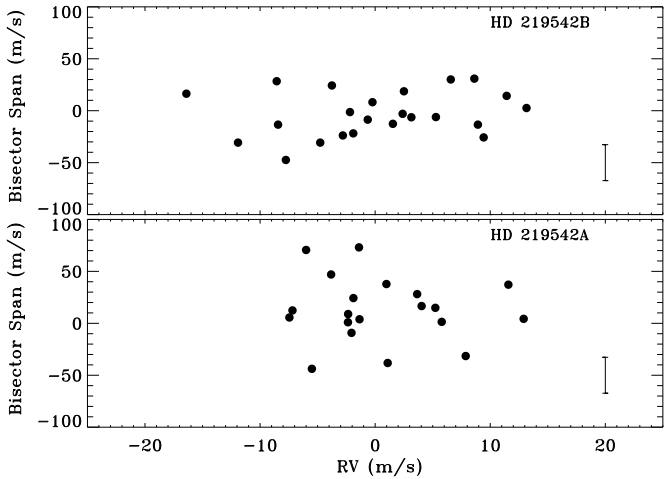}
      \caption{Bisector span measurements vs radial velocities for
               HD~219542~B (upper panel) and HD~219542~A (lower panel).
               There are not significant correlations for both stars.
               On the right corner the typical short term error
               on bisector span measurement is shown.}
         \label{f:bis_vs_rv}
   \end{figure}

In the case of HD~219542, the scatter of bisector spans is 31.2 m/s and
21.8 m/s for A and B component respectively. Bisector span is
not significantly correlated with radial velocities for both stars
(Fig~\ref{f:bis_vs_rv}).
Lower limits to measurement errors, determined as above from
the results of consecutive spectra, is about 17.3 m/s.
Therefore the observed pattern can be explained mostly by measurement
errors. Anyway, there is still room for bisector span variations
corresponding to a radial velocity variations of $\sim 15$ m/s.
This is larger than the observed scatter for HD~219542~B, so that
we can not exclude that the observed radial velocity variations are due to
line asymmetries.

We note that while the  bisector analysis is not conclusive in the 
present case, however it would be sufficient to
disentangle the origin of radial velocity variations with amplitudes
of a few tens of m/s.
Finally, we note that a more detailed approach (that takes into account the
instrumental line profile determined using the iodine lines) could be devised;
however it requires a lengthy procedure and we postpone it to a forthcoming
paper.

\subsection{Limits on photometric variations}
\label{s:hipparcos}

The only photometric data set available suitable for a variability search
is the Hipparcos photometry.
However, it refers to the combined light of A and B components.
When averaging data on a daily basis, the r.m.s. is 0.004 magnitudes,
without hints of variability above the measurement errors.
No significant periodicities appeared applying the same procedures
used in Sect.~\ref{s:statistic}.
The study of the Hyades by Paulson et al. (quoted by 
Henry et al.~\cite{henry02}) 
gives a relation between the radial velocity and photometric scatters of 
$\sim3.3$ m/s/mmag. Therefore, to justify the observed radial
velocity dispersion, a photometric scatter 0.003 mag
would be required.
This is much lower than the limit placed by Hipparcos photometry, taking into
account that HD~219542~B contributes only $\sim$ 40\% of the 
composite light.


\subsection{Equivalent Width Variations}
\label{s:ew}

We expect that activity may cause (small) variations in the average temperature
in the line forming region. For this reason, we expect (small) variations in
the average equivalent widths ($EW$s) of spectral lines. This method is similar
to the line depth ratio method considered by Gray (\cite{gray94}).
We then selected among the line list used in Sect.~\ref{s:bisector} in the
spectrum section clean of iodine lines the lines with large sensitivity 
to effective temperature changes, determined using appropriate 
spectral synthesis.
The final selection includes about 40 lines for which variations of 1.0~m\AA\
in the $EW$ are due to temperature changes of less than 20 K.
A variations of 20 K in the stellar effective temperature 
causes a variation of about 0.017 mag in stellar magnitude.
We then measured the $EW$s of the lines on all spectra of both components of
HD~219542, as well as for the comparison star 51 Peg, using an our own 
purposely written code. On 51 Peg we found that for a single line, the
typical r.m.s.
scatter of the residuals of the $EW$s with respect of the average value 
is 0.86~m\AA. The spectra have a typical S/N of about
170 at this wavelength. Application of the formula by Cayrel (\cite{cayrel})
would produce an expected error of 0.22~m\AA\ for these lines; 
however this formula
neglects the contribution due to the continuum and background subtraction, 
that likely significantly increases the errors. 
Furthermore, internal errors obtained by considering the error
for individual lines divided by the square root of the number of lines,
likely underestimates the real error bar because they do not take into
account correlation existing between nearby lines (due to positioning of the
continuum level and errors in the background subtraction). 
We then measured the temperature variations properly weighting the lines
according to their temperature sensitivity.
The observed r.m.s. scatter of results for individual spectra is 2.5 K,
with internal errors 2.1 K. 
This yields an upper limit to real spectrum-to-spectrum of
1.3~K. This on turn yields an upper
limit to real variations in the luminosity of 51 Peg of 0.0011~mag (comparable
to the results by Henry et al.~\cite{henry00}). This
very strict upper limits constraints the jitter in radial velocity to be 
less than 3.6 m/s (using the quoted relation 3.3 m/s/mmag).

When the same procedure is applied to the spectra of HD~219542~B, we get much
less stringent upper limits, due to the lower S/N of the spectra. The observed
r.m.s. of temperature is 8.34 K; the internal error is
5.95~K. This implies an upper limit to real variations in the effective
temperature  of 5.8~K, and an upper limit to photometric variations
of 0.0050~mag, about twice lower than the limit determined from the
Hipparcos photometry. However, this limit is still large enough to accomodate
a jitter in the radial velocity of about 15~m/s, similar to the
limit obtained from bisectors and well enough to justify 
observations.

We conclude that also this estimate of the expected jitter is not accurate
enough for the present purpose.

\subsection{Expected radial velocity jitter}
\label{s:jitter}

The low chromospheric activity and the low rotational velocities
($R_{HK} = -5.36 \pm 0.13$, $v\sin i =1.9 $~km/s for HD 219542 A;
 $R_{HK} = -5.11 \pm 0.13$, $v\sin i =2.4 $~km/s for HD 219542 B)
indicate a low activity level. 
The jitter expected on the basis of the relation $\sigma_{v}$ vs $v \sin i$ by 
Saar et al.~(\cite{saar}) is about 6 m/s and 8 m/s for HD 219542 A and B
respectively.
The $\log R^{'}_{HK}$ calibration gives instead $\sim 2$ and $3 $~m/s 
respectively.
The calibration by Marcy (Marcy 2002, private communication), that is
based on a larger and higher precision data set than Saar et al.~(\cite{saar})
ones and takes into account the color dependence, predict instead 
a jitter of about 2 and 4 m/s for  components A and B respectively.

To get a jitter of 8 m/s (required to explain the observed scatter),
the chromospheric activity of HD~219542~B should be as high as 
$\log R^{'}_{HK}\sim -4.75$. Such a value seems to be excluded from 
the analysis presented in Sect.~\ref{s:activity_indic}\footnote{Note that 
our $R_{HK}$ measurement was obtained in the season for which a larger 
radial velocity scatter is present.}.

The observed radial velocity jitter of HD~219542~A is very low, at  most 4 m/s,
assuming that the 25\% overestimate of errors of consecutive
spectra holds also on longer time baseline.
On the basis of  the results by Paulson et al.~(\cite{paulson02})
for the Hyades, we do not expect that the colder component of a (coeval) pair
shows larger jitter than the warmer one. 
However, the two stars may be on different phases
of their activity cicle, so that 
the low jitter of HD 219542 A does not automatically 
guarantee the absence of moderate activity in its 
companion\footnote{If the planet around HD~219542~B does exist, 
the upper limit to the jitter 
estimated from residuals from the best fitting orbit, is 4.2 m/s, similar
to HD~219542~A.}.

Therefore, we can not exclude an activity origin
for the velocity variations, taking into account that the line bisector,
photometry and equivalent width analysis do not have the precision 
required to evidence  the variations causing a $\sim 10$ m/s 
radial velocity scatter.

\section{Limits on planetary companions}
\label{s:noplanet_a}

As shown in Sect.~\ref{s:diff_rv}, HD~219542~A does not show variations above
errors and the statistical analysis did not reveal any significant
periodicity in the velocities.
The small velocity errors and the dense sampling of the observations allow us
to derive firm upper limits for possible substellar companions.
To this aim, we have developed a procedure that allow us to  derive
constraints for the masses of planets in circular as well as in 
eccentric orbit.

The method is a Monte--Carlo extraction and is as follows:
we set a dense grid of mass and semimajor axis values
covering the ranges  0.05-4$M_{J}$ and 0.01-3 AU.
For each pair of mass and semimajor axis we randomly generated
(with uniform distribution)
10000 combinations of orbital phase $T_{0}$
eccentricity $e$ and longitude of the periastron $\omega$
(the last two only for the eccentric case).
We adopt the following ranges:
$e<0.5625 \times \log(P)-0.12$ for $P<80$ days (a conservative upper limit 
to the eccentricity of known exoplanets as a function of period) and 
$0 \leq e \leq $ 0.95 for longer periods\footnote{This upper limit
is justified on the basis on the detection of a planet with
$e$=0.93 (HD80606b, Naef et al.~\cite{naef}).};
$0 \leq  \omega \leq  360$ deg; 
$0 \leq T_{0} \leq P_{max}$.

For each set of orbital parameters, we create a synthetic radial velocity
curve obtained by adding a Gaussian noise term
(with the same rms as our RV-results) to a Keplerian orbit sampled at
the epochs of our observations.

We then applied the F test, and computed the probability that
the synthetic and the observed data sets have the same variance. 
If the probability is larger than 5\% for more than 95\% of the 10000
data sets for each mass--semimajor axis pair, 
than we took the planet with the 
corresponding mass and orbital semiaxis as compatible with the 
observed data. 
Otherwise, we considered the planet as excluded by our data, 
since its presence would have created a detectable excess of variability.

\begin{figure}
  \includegraphics[angle=-90,width=9cm]{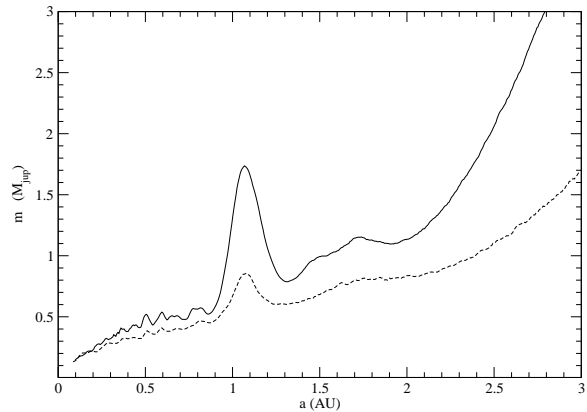}
      \caption{Limits on planets around HD~219542~A in circular (dashed line)
               and eccentric (continuous line) 
               orbits derived using  the Monte Carlo
               procedure described in the text.}
         \label{f:limits}
   \end{figure}

\begin{figure}
   \includegraphics[angle=-90,width=9cm]{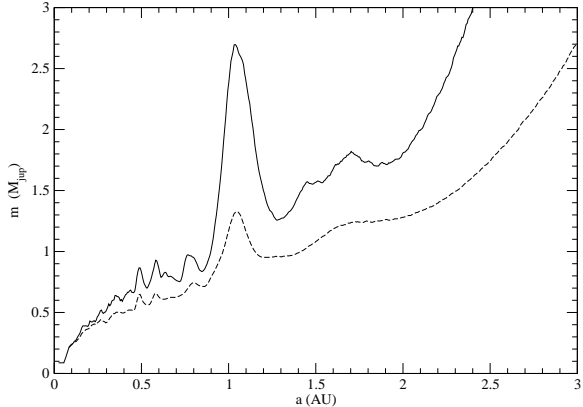}
      \caption{Limits on planets around HD~219542~B in circular (dashed line)
               and eccentric (continuous line) 
               orbits derived using  the Monte Carlo
               procedure described in the text.}
         \label{f:limits_b}
   \end{figure}

In this manner we obtain two regions in the mass--semimajor axis plane
the first where planets can be excluded, and  an area  
below our limits (see Fig.~\ref{f:limits}-\ref{f:limits_b}), where planets
can still exist.
It must be noted that the derived upper limits concern $m \sin i$.

Our method, based on the evaluation of the excess of radial
velocity variability caused
by the presence of hypothetical planets, allows a complete exploration
of the possible orbital parameters for eccentric orbits (the
real case, since most of the known planets are in eccentric orbits).
This is not feasible when the synthetic data are examined using methods like
the periodogram analysis (see Endl et al.~\cite{endl02}).

Note that this procedure attempts to answers the question: 
would the presence of a
planet more massive than this limit at the given separation have caused 
scatter incompatible with observed data?
This is a somewhat different question from that answered by the usual
procedure, that consider the detection of a planet with a high level of
confidence (typically 99\%) using procedures based on the periodogram 
analysis (see e.g. Endl et al.~\cite{endl02}; Cumming et al.~\cite{cumming}).
These authors formulate the problem slightly differently: they
determine the minimum amplitudes for planetary signal at various separations, 
which can be detected by their data at a statistical significance of $>99$\%.
The two approaches are rather different, because in principle, a
planet could be present even if its detection is not significant at a high
level of confidence. However, we repeated our procedure for the case of
$\alpha$\ Cen A, for which upper limits to planet masses have been obtained
by Endl et al.~(\cite{endl01}) using the alternative approach, 
and we find that our procedure (when assuming circular orbits) closely 
reproduces the same upper limits to planetary masses obtained in that paper.

Fig.~\ref{f:limits} shows the results for HD~219542~A.
The limits for planets in eccentric orbit are less stringent than
for planets in circular orbits, becuase of the flexibility of shape 
of the keplerian curve for high eccentricties.
Planets with $m \sin i > 2~M_{J}$ are excluded for separation smaller
than 2.4~AU also 51 Peg-like planets with masses
$m \sin i > 0.15~M_{J}$ are excluded. 
For orbital periods exceeding the baseline of our measurement
the constraint becomes shallower. A peak corresponding to the 1 year observing
window can be noticed, especially in the eccentric case.
Fig.~\ref{f:limits_b} shows the results for HD~219542~B. The limits
are less stringent because of the larger scatter of the velocities.

\section{Discussion}

The main results of this paper are the following:
\begin{enumerate}
\item
We detected a low-amplitude RV variation for HD~219542~B which could
be due to a Saturn mass planet in close orbit;
\item
For component A, the star which shows signs of planetary accretion,
we can place stringent upper
limits on the existence of orbiting giant planets.
\end{enumerate}

Taking into account the possibility that 
that the observed radial velocity variation of component B  is caused 
by modest stellar activity, we think that a lengthy discussion 
on the implications
of the presence of a planet around HD~219542~B is premature.

However, we note that HD~219542~Bb, if it exists, should represent 
the third case of a planet in a binary system around  
the lithium poor component 
of the pair, after 16 Cyg B and HD 178911. In the case of HD 219542
and 16 Cyg\footnote{To our knowledge, a high precision differential 
chemical composition analysis was not performed for HD 178911, but 
this would be complicated by the double line spectroscopic binarity of 
the primary.}, the star without detected planet shows also the signatures 
of accretion of planetary material.
In these cases, the signatures of accretion might be anti-correlated
with the presence of planets.

HD~219542~B is by itself a metal-rich star, with metallicity 
typical for the stars with planets ([Fe/H]=+0.20).
If we take the abundances of HD~219542~B as the original values
one for the system (i.e. assuming that no enrichment occurred for HD~219542~B)
it implies that the system was originally  very metal rich.
This might suggest  that accretion events are more likely on very metal
rich stars, possibly because they form more planets and/or they have
more massive disks. The full analysis of the SARG sample would 
tell us if such enrichment events are limited to very metal rich stars
or they can happen also for metal poor stars.

The lack of giant planets in close orbit around HD~219542~A leaves 
several possibilities for the origin of the observed signatures of accretion
open.

\begin{itemize}
\item 
Planet formation did not occur around HD~219542~A and a significant
portion of the circumstellar disk fell into the stars possibly
because of the dynamical perturbation of HD~219542~B. This possibility
is neither excluded nor supported by the possible stellar 
orbital parameters given in Sect.~\ref{s:bin_orbit}.
If the planet around component B is confirmed, it would 
indicate that gravitational influence did not disrupt the planet 
formation process.
\item
One or more giant planets orbiting around HD 219542 A were scattered toward the
star by other planets in the system (Marzari \& Weidenschilling 
\cite{marzari02}) or by the perturbation of HD 219542 B.
In the first case, we would expect to find a planet in an external eccentric
orbit. The continuation of our radial velocity monitoring could reveal
that.
\item
No giant planets formed in the system  while terrestrial ones formed in the 
inner regions. Some of them
may have been scattered into the star by mutual interactions.
\end{itemize}

Our results exclude instead a  scenario with 
a giant planet that formed near the ice boundary (4-5 AU) and migrated 
inward, causing the infall on the star of the inner part of the disk
and/or terrestrial planets already formed (we would have detected
such planet, unless the planet itself was engulfed by the star).

The study of a few pairs clearly does not allow us to reach a
firm conclusion about the frequency of pairs with chemical composition 
differences; a general picture is missing.
How frequent are these pairs?
Is there any link between the pollution phenomena and some properties like 
the physical separation of the components and  the initial
composition?
How frequent are planets around the components of nearly equal mass binaries
with separation 200-300 AU and what are their orbital characteristics?
Are planets really missing around polluted stars?

Our on-going survey including a sample of about 50 binaries
might allow some deeper insight into these very exciting topics.

\begin{acknowledgements}

   We warmly thank  A.~Bragaglia,  F.~Leone,  S.~Randich, F.~Sabbadin, and 
   P.~Sestito for taking some spectra during their observing time.
   We thank the TNG staff for its help in the observations.
   We are grateful to R.E.M.~Griffin for supplying the {\it
   Orbitsolver} code. We are grateful to M.~K\"urster and S.~Els
   for useful insights on the radial velocity measurements and to
   G.~Marcy for providing us its jitter calibration.
   We thank the referee A.~Hatzes for a very encouraging report.
   This publication makes use of public data obtained during
   FEROS Commissioning and of data products from the Two Micron All Sky Survey,
   which is a joint project of the University of Massachusetts and the
   Infrared Processing and Analysis Center/California Institute of Technology,
   funded by the National Aeronautics and Space Administration and the
   National Science Foundation. 
   This research has made use of the  SIMBAD database, operated at CDS, 
   Strasbourg, France.
   ME acknowledges support from NSF Grant AST-9808980 and NASA Grant NAG5-9227.

\end{acknowledgements}

\end{document}